\documentclass[aps,preprint,showpacs,showkeys,groupedaddress]{revtex4}  
\usepackage{graphicx}  
\usepackage{dcolumn}   
\usepackage{bm}        
\usepackage{amssymb}   
\usepackage{subfigure}
\usepackage{float}
\usepackage{amsmath,amsfonts}
\usepackage{setspace}
\usepackage[utf8]{inputenc}

\begin{document}

\title{Complex free energy landscapes in biaxial nematics and role of 
repulsive interactions : A Wang - Landau study}

\author{B. Kamala Latha$^{1}$}

\author{K. P. N. Murthy$^{1}$} 

\altaffiliation{Present address: Manipal Centre for Natural Sciences, Manipal University, Manipal 576104, India}

\author{ V. S. S. Sastry$^{1,2}$}
\affiliation {$^{1}$School of Physics,University of Hyderabad, Hyderabad 500046, India}
\affiliation{$^{2}$Centre for Modelling, Simulation and Design, University of 
Hyderabad, Hyderabad 500046, India}

\date{\today}
\begin{abstract}

General quadratic Hamiltonian models, describing interaction between 
crystal molecules (typically with $D_{2h}$ symmetry) take into account
couplings between their uniaxial and biaxial tensors. While the 
attractive contributions arising from interactions between similar
tensors of the participating molecules provide for eventual condensation
of the respective orders at suitably low temperatures, the role of 
cross-coupling between unlike tensors is not fully appreciated. Our
recent study with an advanced Monte Carlo technique (entropic sampling)
showed clearly the increasing relevance of this cross term in determining
the phase diagram, contravening in some regions of model parameter
space, the predictions of mean field theory and standard Monte Carlo
simulation results. In this context, we investigated the phase diagrams
and the nature of the phases therein, on two trajectories in the parameter
space: one is a line in the interior region of biaxial stability believed
to be representative  of the real systems, and the second is the
extensively investigated parabolic path resulting from the London 
dispersion approximation. In both the cases, we find the destabilizing
effect of increased cross-coupling interactions, which invariably result in the 
formation of local biaxial organizations inhomogeneously distributed.
This manifests as a small, but unmistakable, contribution of biaxial 
order in the uniaxial phase.The free energy profiles computed 
in the present study as a function of the two dominant order parameters
indicate complex landscapes. On the one hand these profiles account
for the unusual thermal behaviour of the biaxial order parameter under
significant destabilizing influence from the cross terms. On the other,
they also allude to the possibility that in real systems these complexities
might be indeed inhibiting the  formation of a low temperature biaxial 
order itself - perhaps reflecting the difficulties in their ready 
realization in the laboratory.  
 \end{abstract}            

\keywords{Biaxial liquid crystals, entropic sampling, Wang-Landau 
algorithm, Free energy computations}
\pacs{64.70.M-,64.70.mf}                             
\maketitle
            
\section{Introduction}
The biaxial nematic phase, proposed initially in the work by Freiser and Straley 
\cite{Freiser, Straley}, has been the subject of many theoretical 
\cite{ Alben,Luck75, Bocara, Gramsbergen, Remler, Mulder, Teixeira} and 
experimental \cite{Acharya04, Madsen, Merkel, Figueirinhas05, Severing} 
investigations in the recent years, and was investigated extensively 
by Monte Carlo (MC) simulations 
\cite{Luckhurst80, Allen90, Biscarini95, Camp97, Chiccoli99, romano,
romano1, Preeti11, Kamala14, Kamala15}. However their experimental 
realization was not so readily possible  and is still a matter of  debate 
 \cite{Van Le, Vaupotic, Mamatha}.

According to recent mean field (MF) treatments \cite{Sonnet, Longa05, 
 matteis05A, matteis05B, Bisi06,Dematteis07, Dematteis08}, the relevant 
 Hamiltonian parameter space conducive to the formation of stable 
 biaxial phase comprises of a triangular region (say, $\Delta$) in the 
 $(\gamma, \lambda)$ plane (the  essential triangle) shown in 
 Fig.~\ref{fig:1} \cite{Bisi06}, the long axes of the molecules defining
 the primary director. The quadratic Hamiltonian for the biaxial system 
 adds to the dominant attractive interaction between the major molecular 
 axes of the neighbouring molecules (i.e Lebwohl - Lasher interaction
  term \cite{LL}) two more terms: a coupling between the two molecular 
 biaxial tensors with strength $\lambda$ and a cross-coupling between 
 the biaxial and uniaxial tensors of the two molecules, through $\gamma$. 
 The  MF predictions and our earlier MC work \cite{Kamala14, Kamala15}, 
focussed on two specific paths in this plane which have axial 
symmetry of the torques: along the $\lambda$ - axis ($D_{4h}$ symmetry 
of molecular pairwise interactions around the molecular z-axes), and 
the diagonal IV (with similar symmetry around the molecular x-axes), 
see Fig.~\ref{fig:1}. The deviations from 
the MF work became discernible when $\gamma$ is appreciable, with the 
corresponding interactions competing with those of $\lambda$, along the 
path $IV$ (Fig.~\ref{fig:1}). Earlier MC simulations based on standard 
Metropolis sampling methods \cite{Dematteis08}, while
 being generally supportive of MF results, were qualitatively deviating 
from our MC data, obtained through entropic sampling methods. We reported
an additional intermediate biaxial phase in the MF predicted 
direct $(N_{B} - I)$ transition sequence, starting from the point $K$
 and extending upto the point $V$ encompassing the Landau point 
 $\textit{T}$ \cite{Kamala15}. $\textit{T }$ is special since 
it represents a pure biaxial interaction between the two major axes 
(y and z) with  $D_{4h}$ pair-interaction symmetry and with no uniaxial
 coupling between the minor axes (x-axes). Incidentally it also 
 represents a cross-over point on the dispersion parabola ($OT$) from 
 the prolate to oblate molecular symmetry.  

     The more realistic choices for $(\gamma, \lambda)$ values appear 
more likely to be within the $\Delta$ region as has been reported 
experimentally  recently \cite{Bisi08}. Also of particular interest
in the earlier literature are models which correspond to systems 
satisfying the London dispersion approximation \cite{Luck75,
Luckhurst80}, reducing the number of independent model parameters
 to one, with $\lambda = \gamma^{2}$.
Phase diagram along parabolic trajectory has been extensively studied
earlier \cite{Biscarini95,Chiccoli99}, and has been used as a prototype 
for several investigations \cite{Berardi03, Chiccoli1, Chiccoli2, Berardi08,
BerardiB}. The dispersion parabola also defines an 
interesting boundary separating regions of ($\gamma, \lambda$) parameter
 space: one region that makes the Hamiltonian fully attractive above
 the parabola and the other which makes it partly repulsive  (below the
 parabola) \cite{Bisi06}. Investigation on the nature of the phases
 with entropic sampling techniques as one traverses the parabola from
 Lebwohl-Lasher limit (origin) to the Landau point $\textit{T}$, could
 be interesting from the point of understanding the destabilizing
 influence of $\gamma$ along this path, if any.
 
      In this work, we carried out a detailed entropic sampling based 
 MC study of the phase diagram on a straight line path within the 
 triangle (IW in Fig.~\ref{fig:1}), where W is the 
 mid point of OV.  The relative importance of the cross-coupling $\gamma$
 term increases along the path IW which intersects the parabola at point C, 
  beyond which the $\gamma$ -term provides a repulsive contribution to 
  the Hamiltonian. We supplement 
this data with results from standard Boltzmann ensembles for comparison. 
With the density of states obtained from the entropic method, we 
compute the free energy profiles as functions of order parameters, with
a view to correlating them with the observed thermal behaviour of these
variables. A similar study was carried out at several points
on the parabola. It is interesting to observe the curious changes that
the model induces on the  macroscopic behaviour, as it starts with a 
small perturbation on the LL-model near the origin and moves
all the way to the Landau point $\textit{T}$. This paper discusses the 
MC results along these two trajectories.  

The paper is divided into five sections. The mean field Hamiltonian model
and its representation for purposes of simulation are 
outlined in section II. The details of entropic sampling based simulation 
are described  in section III. The results are presented and discussed 
in section IV. We summarise the salient features of the work in section V.
  
\section{Hamiltonian Model}        
   The general interaction between two liquid crystal molecules with 
 $D_{2h}$ symmetry limited to quadratic terms, each described by two 
 symmetric traceless tensors ($\bm{q}$, $\bm{b}$) and ($\bm{q^{'}}$, 
 $\bm{b^{'}}$), is expanded as  $ H=-U[\xi \, \bm{q} \cdot \bm{q}^{\, \prime}
+ \gamma(\bm{q} \cdot \bm{b}^{\, \prime} + \bm{q^}{\, \prime} \cdot \bm{b}) +
\lambda \, \bm{b} \cdot \bm{b}^{\, \prime}]$. Here $\bm{q}$ and $\bm{b}$
are the irreducible components of the anisotropic parts of the molecular
susceptibility tensor, and can be represented in its eigen frame 
$(\bm{e},\bm{e_{\perp}},\bm{m})$ as $\bm{q} = \bm{m} \otimes \bm{m} -
\frac{\bm{I}}{3}$ and $\bm{b} = \bm{e} \otimes \bm{e} - \bm{e}_{\perp}
\otimes \bm{e}_{\perp}$. Similar representation (for a neighboring
molecule)  holds for $\bm{q^{'}}$, $\bm{b^{'}}$ in the eigen frame 
$(\bm{e^{'}}, \bm{e^{'}_{\perp}}, \bm{m^{'}})$, following the 
notation used earlier \cite{Sonnet}. MF
analysis of the Hamiltonian predicts a triangular region OIV 
(Fig.~\ref{fig:1}) \cite{Bisi06} as the region of stability for
the biaxial phase in  the interaction parameter $(\gamma, 
\lambda)$ space, assigning the primary director to the orientation of 
 the long molecular axes. The dispersion  parabola ($\lambda = 
 \gamma^{2}$) OCT divides the parameter space into 
 two regions  - the region above within the triangle, OIT - where the
 interaction Hamiltonian is globally attractive, and the one below 
 OTV - where the interaction is partly repulsive due to the $\gamma$-term. 
 The points $C_{1}$  and $C_{3}$ are tricritical points of the 
uniaxial  nematic - biaxial nematic transition and $C_{2}$ is a triple 
point hosting the three phases of the medium: isotropic (I), uniaxial 
nematic $(N_{U})$ and biaxial nematic $(N_{B})$ phases \cite{Bisi06}. 
(K is a point where  the $N_{B} - I$  phase sequence has been found to 
change to  $N_{B} - N_{B_{1}} - I $ \cite{Kamala14}, deviating from 
the MF prediction). MF also predicts
 a direct $N_{B} - I$ transition inside the parameter region $IC_{2}C_{3}$,
 and tricritical nature for $N_{U} - N_{B}$ transition along $C_{1}C_{3}$
 \cite{Bisi06}. The MF analysis based on mini-max principle
 involving only the two dominant order parameters (out of the four)
  permits the existence of a biaxial phase even at 
 the base point V of triangle ($\lambda$ = 0), though such a phase is
  forbidden on grounds of  biaxial stability \cite{Dematteis07}.
 
            For the purpose of  simulations, the mean field Hamiltonian 
 is conveniently recast in terms of a  biaxial mesogenic  lattice
 model, where two molecules of $D_{2h}$ symmetry at distinct lattice 
 sites, represented by orthonormal triplet of 3-component unit vectors
  $\bm{u}_{a}$, $\bm{v}_{b}$ (a,b = 1,2,3) interact through 
a nearest neighbour pair potential \cite{romano}
\begin{equation}
U= -\epsilon \lbrace G_{33} - 2\gamma(G_{11}-G_{22})+
\lambda[2(G_{11}+G_{22})-G_{33}]\rbrace.
\label{eqn:w1}
\end{equation}
Here  $f_{ab}$= ($\bm{u}_a$.$\bm{v}_b$),
 $G_{ab}$=$P_2$($f_{ab}$), with $P_{2}$ denoting the second Legendre 
polynomial. $\epsilon$ is  a positive quantity setting the reduced temperature 
$\textit{T}^{'}=k_{B}\textit{T}/\epsilon $, where $\textit{T}$ is the 
absolute temperature the system. In these simulations $\epsilon$ is set
to unity.

 \section{Details of simulation}
  The Wang-Landau (WL) algorithm \cite{Wang}, addresses the problem of 
 efficient entropic sampling of the configuration space to construct 
 ensembles (in discrete spin systems), that are uniformly distributed 
 with respect to energy, through an accurate estimation of the density
 of states (DoS) of the system. This has been successful in tackling 
 various problems in statistical physics \cite{Landau1,  Murthy1} 
 as diverse as  polymers and protein folding \cite{Rathore03, Seaton10,
 Priya11}, self assembly \cite{Landau13} and is being continually 
 updated for application to more  complex systems
  \cite{Poulain,Sinha09,Raj,Yang13, Vogel13, Katie14, Xie14}. 
 The algorithm was suitably modified for application to systems with continuous 
 degrees of freedom like liquid crystals \cite{jayasri05}, and this
 procedure is further augmented with frontier sampling technique
 \cite{Zhou, Jayasri} in order to simulate the bulk  biaxial liquid 
 crystal \cite{Kamala15}. The WL simulation \cite{Wang} estimates the
 DoS, while updating a trial density  $g(E)$ iteratively 
 by  performing a random walk in the energy space with a probability
 proportional to the inverse of the instantaneous $g(E)$,  until a flat
 histogram of energy is achieved as the updation of $g(E)$ is gradually
 withdrawn. The frontier sampling technique introduces additional
  algorithmic guidance to the WL routine, so that lower entropic regions
 are more efficiently accessed. An entropic ensemble of microstates, 
 collected by a random walk guided by the well converged DoS, is fairly 
 uniformly distributed over the energy region of interest, and is 
 adequate to calculate the required thermodynamic properties at the 
 desired temperature resolution by constructing equilibrium
 canonical ensembles (say, RW ensembles) through a reweighting 
 procedure. The free energy profiles, obtained as a function of energy
 and the system order parameters, using the computed DoS, 
 provide further physical insight. We employ here this modified 
 algorithm, described in detail elsewhere \cite{Kamala15}.

 \begin{figure}[]
\centering
\includegraphics[width=0.5\textwidth]{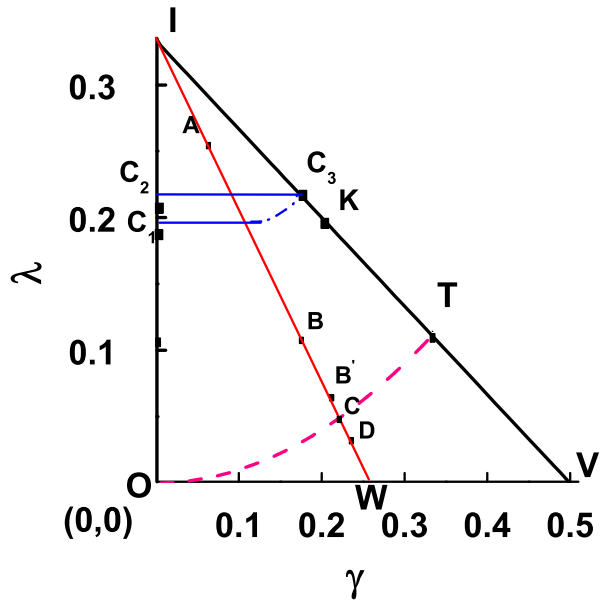}
\caption{(color online) Essential triangle : Region of biaxial stability. 
OI and IV are uniaxial torque lines intersecting at the point I ($0,1/3$). 
OCT is the dispersion parabola which meets the line IV at the Landau point 
T. Base OV is the limit of biaxial stability for the interaction. 
$C_{1}$ and $C_{3}$ are tricritical points and $C_{2}$ is a triple point 
(G. De Matteis $et  \ al$, \textit{Continuum Mech. Thermodyn.} 
\textbf{19} 1-23 (2007)). IW  is the trajectory along which the present  
simulations have been carried out. Points A (0.048, 0.269), B (0.166, 0.111),
$B^{'}$(0.204, 0.061), C(0.215, 0.047) and D (0.225, 0.033) are points of 
particular interest (see text).} 
\label{fig:1}
\end{figure}
     We consider a cubic lattice (size: $L \times L \times L, L= 15, 20$) 
 with periodic boundary conditions. The biaxial molecules 
 on each lattice site interact through the nearest neighbour interaction
 potential in Eqn.~\eqref{eqn:w1}. The parameters $\gamma$ and $\lambda$ are
 chosen such that we traverse  along a trajectory IW passing 
through the apex I and bisecting  the base OV at point W 
(Fig.~\ref{fig:1}). The uniaxial - biaxial coupling 
coefficient $\gamma$ on IW is  half of the the value on the 
diagonal IV, for identical $\lambda$ values. We denote the arclength
of the path OIW as $\lambda^{'}$, given by $\lambda^{'} = \lambda$ on 
segment OI, and $\lambda^{'}= \dfrac{1}{3}(1 + 5\gamma)$ where
$\gamma= \dfrac{(1-3\lambda)}{4}$ on the segment IW.
As we traverse along the trajectory IW  the arc length $\lambda^{'}$ 
varies from $\dfrac{1}{3}$ at I to 0.75 at W. The simulations are done at 
various values of $\lambda^{'}$ on the path IW, starting from the point 
I ($\lambda^{'}$ = 1/3 at $\gamma$ = 0.0, $\lambda$ = 1/3 ) and ending 
at W ( $\lambda^{'}$ = 0.75 at $\gamma$ = 0.25, $\lambda$ = 0.0). Points
A ($\lambda^{'}$ = 0.414 at $\gamma$ = 0.048, $\lambda$ = 0.269), B
 ($\lambda^{'}$ = 0.610  at $\gamma$ = 0.166, $\lambda$ = 0.111),
 $B^{'}$ ($\lambda^{'}$ = 0.674 at $\gamma$ = 0.204, $\lambda$ = 0.061) 
lie in the attractive region for the Hamiltonian, while C ($\lambda^{'}$
 = 0.692 at $\gamma$ = 0.215, $\lambda$ = 0.047) lies on the dispersion 
 parabola and D ($\lambda^{'}$ = 0.709 at $\gamma$ = 0.225, $\lambda$ 
 = 0.033) lies in the partly repulsive region, below the parabola. 

We start the simulation at a chosen value of $\lambda^{'}$  with a 
random orientation of spins on the lattice and the corresponding
 values of  $(\gamma, \lambda)$ are inserted in Eqn.\ref{eqn:w1}
 for calculating the system energy.  An entropic ensemble  comprising 
 of ($ \sim 4 \times 10^{7}$) microstates  is constructed using the 
 Wang - Landau algorithm, with a fairly
 uniform distribution of energy at the end of the simulation. 
Using the DoS computed, canonical ensembles are extracted with reweighting
procedure \cite{Swensden,Berg} at the chosen temperatures (RW-ensembles).
Average values of physical variables are then computed at the required 
temperature resolution. Information on the system energy and the DoS 
facilitate the determination of free energy as a function of energy as 
well as order parameters, at different temperatures.

     Conventional MC simulations based on Metropolis algorithm
(Boltzmann sampling) leading to equilibrium canonical ensembles 
(B-ensembles)  were also carried out at  chosen points on the trajectory 
IW in order to compare the results from both the simulation methods. 
These ensembles were collected, after equilibration, with production 
runs of typically $6 \times 10^{5}$ MC lattice sweeps (Monte Carlo
steps). Temperature variation of the equilibrium averages from B- and 
RW-ensembles are compared, to assess the efficacy of the respective
sampling in the presence of curious free energy terrains in the configuration
space offered by the biaxial system.  

The physical observables of interest, calculated at each value of 
$\lambda^{'}$, are the average energy $<E>$, specific heat $<C_{v}>$, 
energy cumulant $V_{4}$ (= $1-<E^{4}>/(3<E^{2}>^{2})$) 
 which is a measure of the kurtosis \cite{Binder}, the four order parameters of 
the phase calculated according to \cite{Biscarini95,Robert} and their 
susceptibilities. These are the uniaxial order $<R^{2}_{00}>$ (along the 
primary director), the phase biaxiality $<R^{2}_{20}>$, the molecular 
contribution to the biaxiality of the medium  $<R^{2}_{22}>$, and the 
contribution to uniaxial order from the molecular minor axes $<R^{2}_{02}>$.  
The averages are computed at a temperature resolution of 0.002 in the 
temperature ($T^{'}$) range of interest [2.05, 0.05], all in reduced 
units. The error bars for the observables were estimated after 
minimising possible correlations using Jack knife method \cite{Newman1}.
The relative errors in the averages of  energies are found to 
be typically one part in $10^{5}$,  while those in the estimation of 
the averages of the order parameters are 1 in $10^{4}$. 
 
\section{Results and Discussion} 

 We carried out a detailed simulation study employing the entropic 
 sampling  technique, at 30 values of $\lambda^{'}$ with a view to 
obtaining a generic phase diagram inside $\Delta$ along IW. We examined
the temperature dependence of the $C_{v}$ profiles, of the two dominant 
order parameters $<R^{2}_{00}>$ and $<R^{2}_{22}>$, and the Binder's 
energy cumulant $V_{4}$ to determine the phase transition temperatures
and identify the phases. 
\begin{figure}[htp]
\centering{
\subfigure[]{\includegraphics[width=0.35\textwidth]{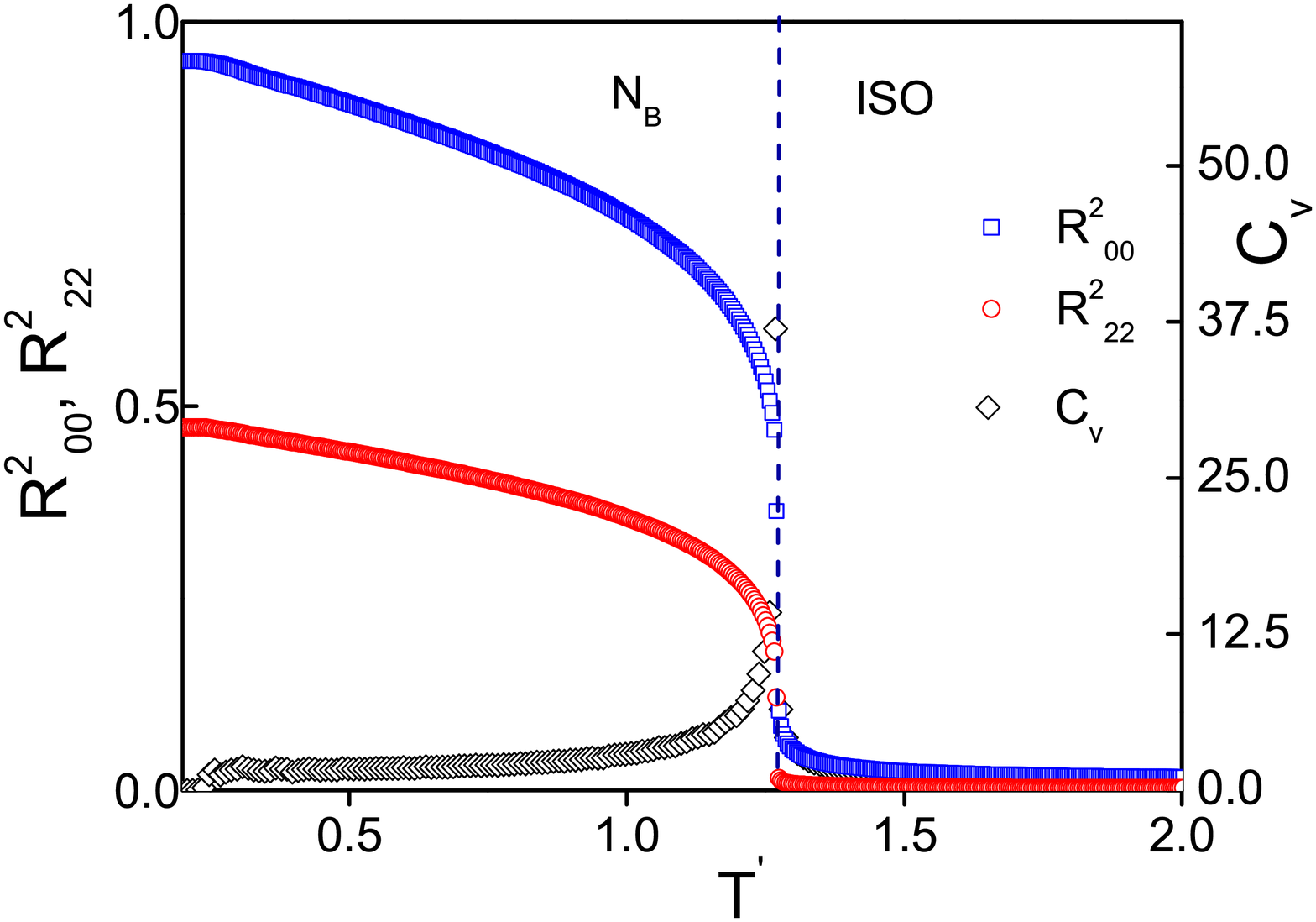}
\label{fig:2a}}
\subfigure[]{\includegraphics[width=0.35\textwidth]{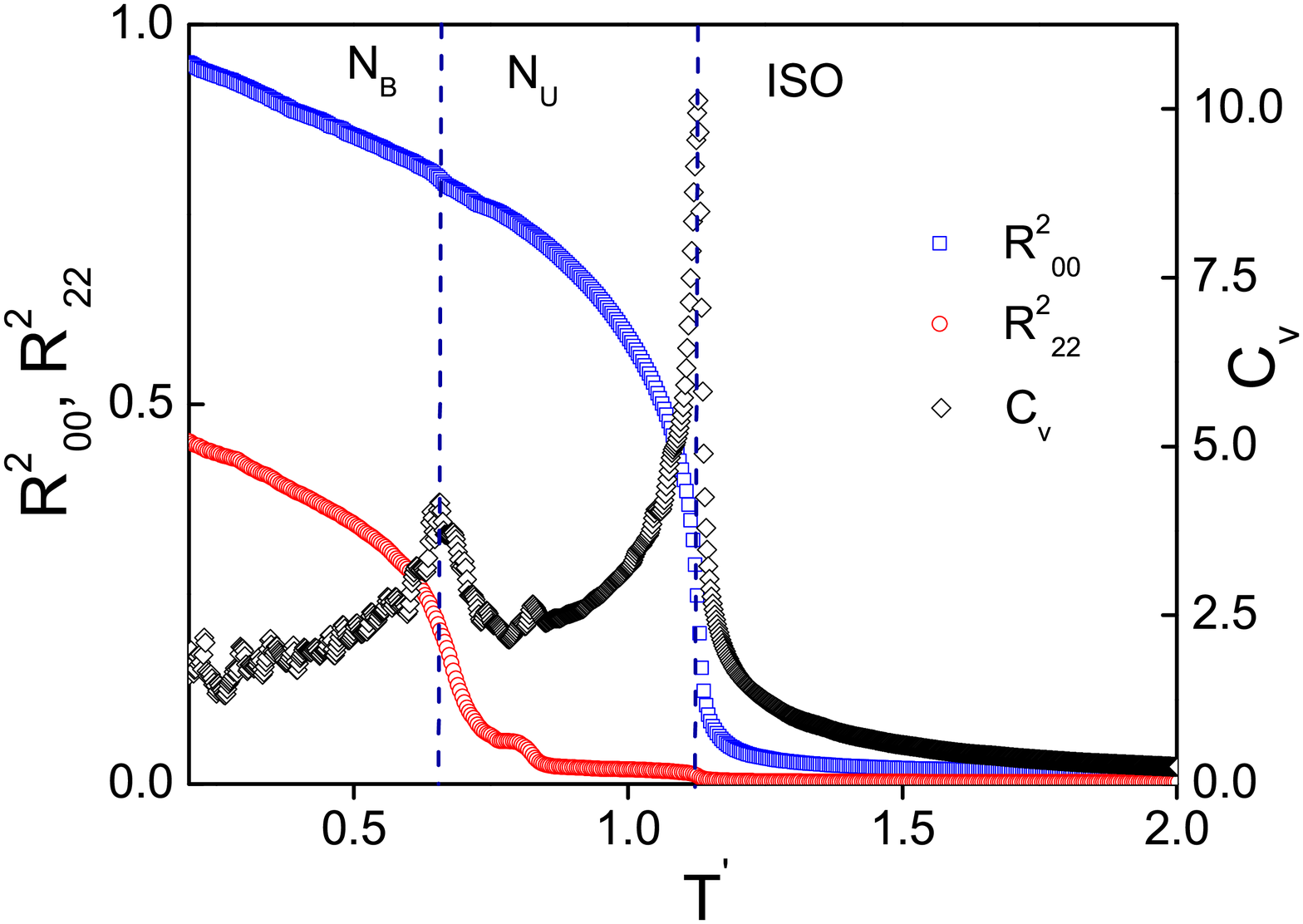}
\label{fig:2b}}
\subfigure[]{\includegraphics[width=0.35\textwidth]{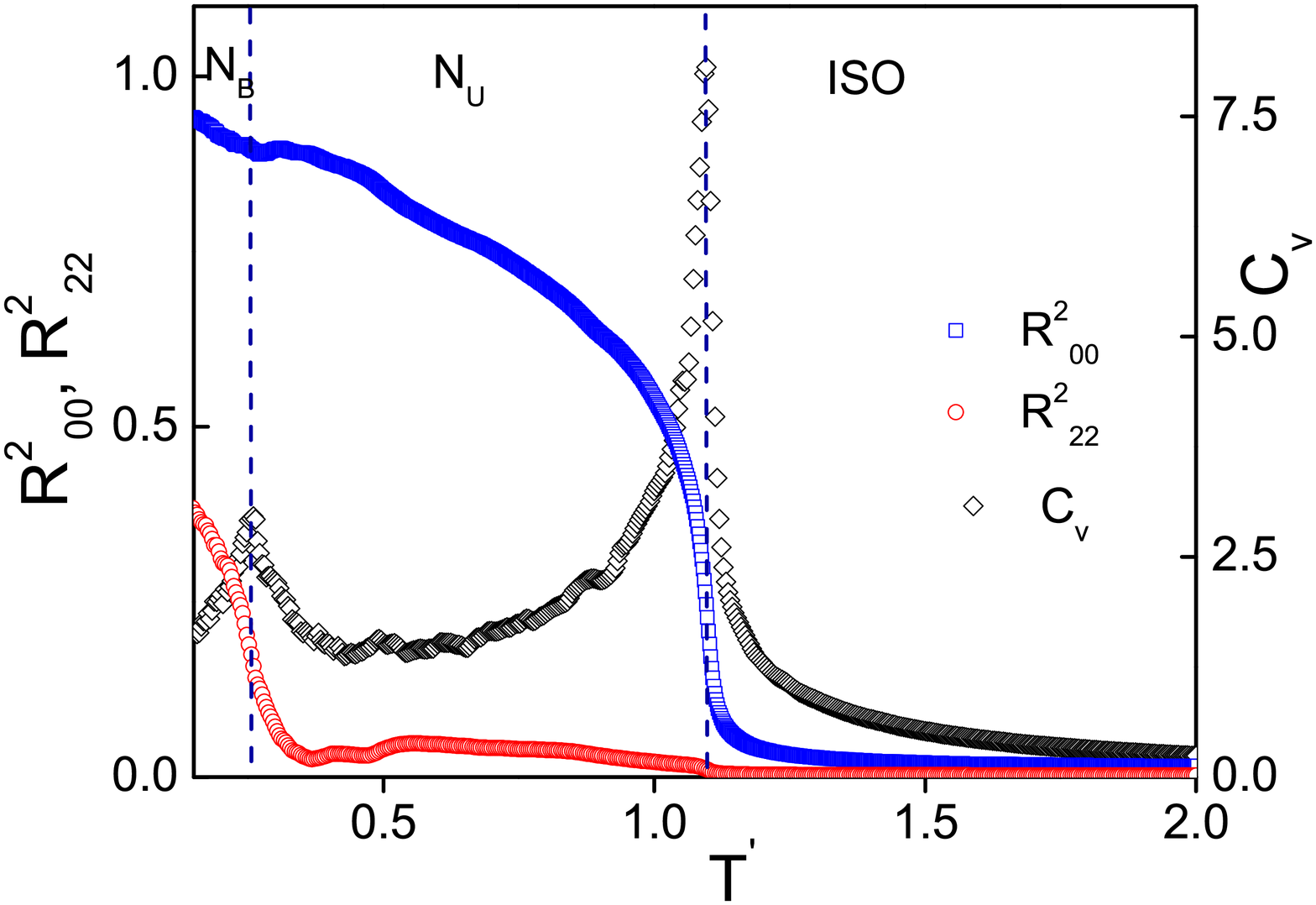}
\label{fig:2c}}
\subfigure[]{\includegraphics[width=0.35\textwidth]{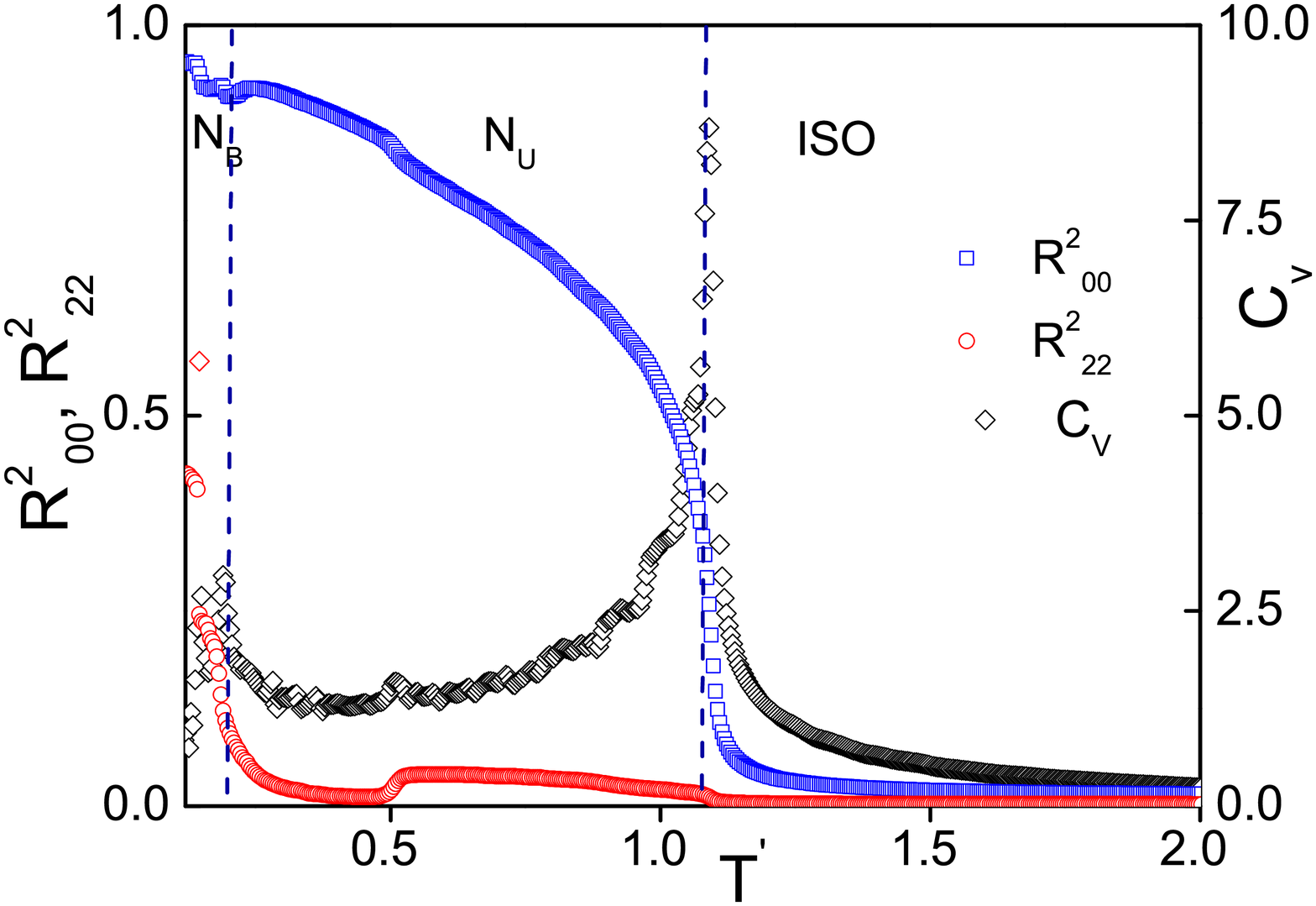}
\label{fig:2d}}
}
\caption{(color online) The temperature variation of order parameters 
superimposed on the specific heat curves to illustrate the phase behaviour 
at the four representative point A,B,C and D with values of  $\lambda^{'}$
= (a) 0.414 (Point A) (b) 0.610 (Point B) (c) 0.692 (Point C) and 
(d) 0.709 (Point D) (L=20)}
\label{fig:2}
\end{figure} 
Typical such data at four representative points,
at A and B in the fully attractive region, at C on the parabola
and at D in the partly repulsive region (Fig.~\ref{fig:1}), are 
presented in Fig.~\ref{fig:2}, with a system size L=20 for point A,B,C
and L=15 for point D.

The order profiles superimposed on the specific heat peaks shown in 
Fig.~\ref{fig:2} indicate that a direct isotropic - biaxial phase 
transition occurs at the point A (Fig.~\ref{fig:2a}). At all
other points two specific heat peaks are observed on cooling, at 
temperatures $T_{1}$ and $T_{2}$. As the system is cooled from the 
isotropic phase, the uniaxial order $R^{2}_{00}$ shows a sharp increase
at $T_{1}$ indicating the onset of an intermediate uniaxial phase 
phase $N_{U}$. It is of interest to note that this  intermediate phase 
also exhibits a small amount of biaxial order which increases to a 
value of $\sim$ 0.03, together with the expected significant increase in the 
uniaxial order as the temperature decreases. The magnitude of $R^{2}_{22}$ 
in this phase seems to be independent of the $\lambda^{'}$ value in the 
attractive region (i.e of the path from I upto C). 
However on the bordering trajectory between the two distinct regions
of the Hamiltonian (i.e point C) and in the partly repulsive region 
(point D) $R^{2}_{22}$ value actually dips on cooling in this intermediate 
phase after the initial onset (Figs.~\ref{fig:2c}, \ref{fig:2d}). The 
$R^{2}_{22}$ value then increases rapidly at the second transition (at $T_{2}$)
 for all values of $\lambda^{'}$, signaling the onset of a low 
 temperature biaxial phase $N_{B}$.
 
\begin{figure}
\centering
\includegraphics[width=0.5\textwidth]{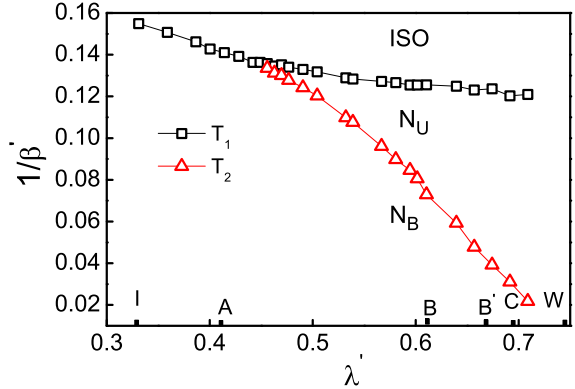}
\caption{(color online) Phase diagram inside the essential triangle 
along path IW} 
\label{fig:3}
\end{figure}

 Based on such study along the trajectory IW
we obtain the phase diagram, as a function of 
the arc length $\lambda^{'}$, shown in Fig.~\ref{fig:3}.
The actual temperature $\textit{T}^{'}$ of the simulation needed to be 
scaled by a factor of 9 (for direct comparison with the Lebwohl-Lasher
(LL) model \cite{LL}), as $\dfrac{1}{\beta^{'}} = \dfrac{T^{'}}{9}$. 

 Beyond the value of 0.709 of $\lambda^{'}$ the parameter region 
 presents dominant cross-coupling term, inducing significant repulsive 
 interactions between the uniaxial and biaxial molecular terms of the 
 neighbouring molecules, thus frustrating the system to find a single, 
stable and  unique free energy minimum. In this scenario, the computational 
times for the convergence of DoS were found to  be impractical at this 
size (L = 20). We are thus constrained to report data in this region 
at L=15. 
 
 We note from the phase diagram (Fig.~\ref{fig:3}) that the biaxial 
 medium undergoes a direct $I - N_{B}$ phase transition for $\lambda^{'}$ 
 values in the  range $1/3$ to 0.455. Thereafter, two 
transitions were observed in the  $\lambda^{'}$ range 0.462 to 0.709. 
The system undergoes a  $I - N_{U}$ transition at the high temperature
 $T_{1}$, followed by a $N_{U} - N_{B}$ transition at the lower temperature
 $T_{2}$. It may be seen that the second transition occurs at 
 progressively lower values of $T_{2}$ which approaches zero asymptotically 
 as the point W (on the base OV) is reached, in conformity with the 
 previous MC simulation results in this limit of $\lambda\rightarrow0$  
  \cite{Luckhurst80}.
     
 \begin{figure}[H]
\centering
\includegraphics[width=0.5\textwidth]{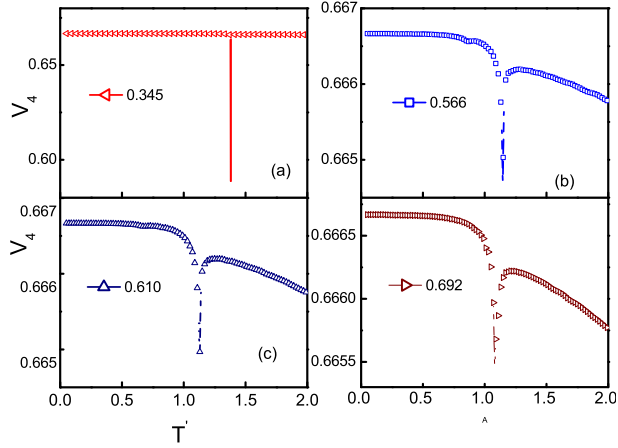}
\caption{(color online) Energy cumulant $V_{4}$ for certain values 
of $\lambda^{'}$ in the range 0.345 to 0.692 (point C)} 
\label{fig:4}
\end{figure}
\begin{figure}[H]
\centering
\includegraphics[width=0.7\textwidth]{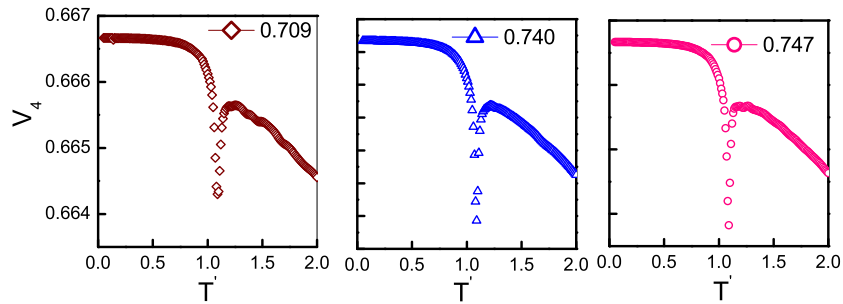}
\caption{(color online) Energy cumulant $V_{4}$ for values of  
$\lambda^{'}$ in the range 0.709 (Point D) to  0.747 at(L=15)} 
\label{fig:5}
\end{figure}

      The nature of the transitions  can be gleaned 
from the plots of the fourth order energy cumulant ($V_{4}$) data 
 shown in Fig.~\ref{fig:4} for some typical values of $\lambda^{'}$. 
The sharp dip in the cumulant value shown in Fig.~\ref{fig:4}(a) at 
$\lambda^{'}$ = 0.345 is indicative of a strong first-order nature 
of the $N_{B} - I$ transition in the $\lambda^{'}$ range 0.345 to 0.45. 
Figs.~\ref{fig:4}(b) - \ref{fig:4}(d) depict the nature of the two 
transitions in the  range of $\lambda^{'}$ from 0.463 to 0.692. The 
$I - N_{U}$ transition at $T_{1}$ shows a progressively weakening first 
order nature (relative to Fig.~\ref{fig:4}(a)), whereas $ N_{U} - N_{B}$ 
transition seems to be continuous over the trajectory. The trajectory
 in the repulsive region also shows similar nature of the transitions
 (Fig.~\ref{fig:5}).
 
\begin{figure}
\centering{
\subfigure[]{\includegraphics[width=0.35\textwidth]{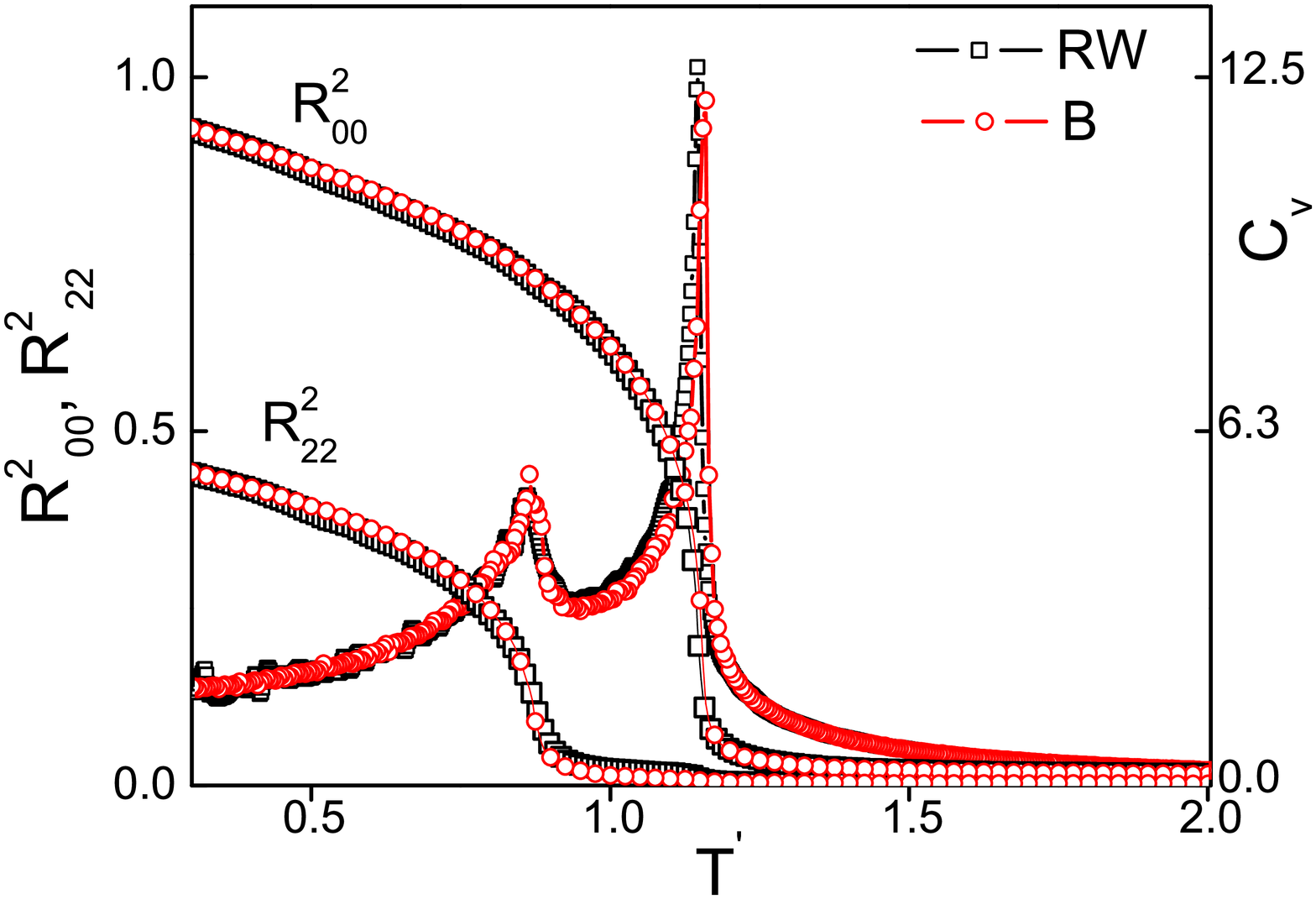}
\label{fig:6a}}
\subfigure[]{\includegraphics[width=0.35\textwidth]{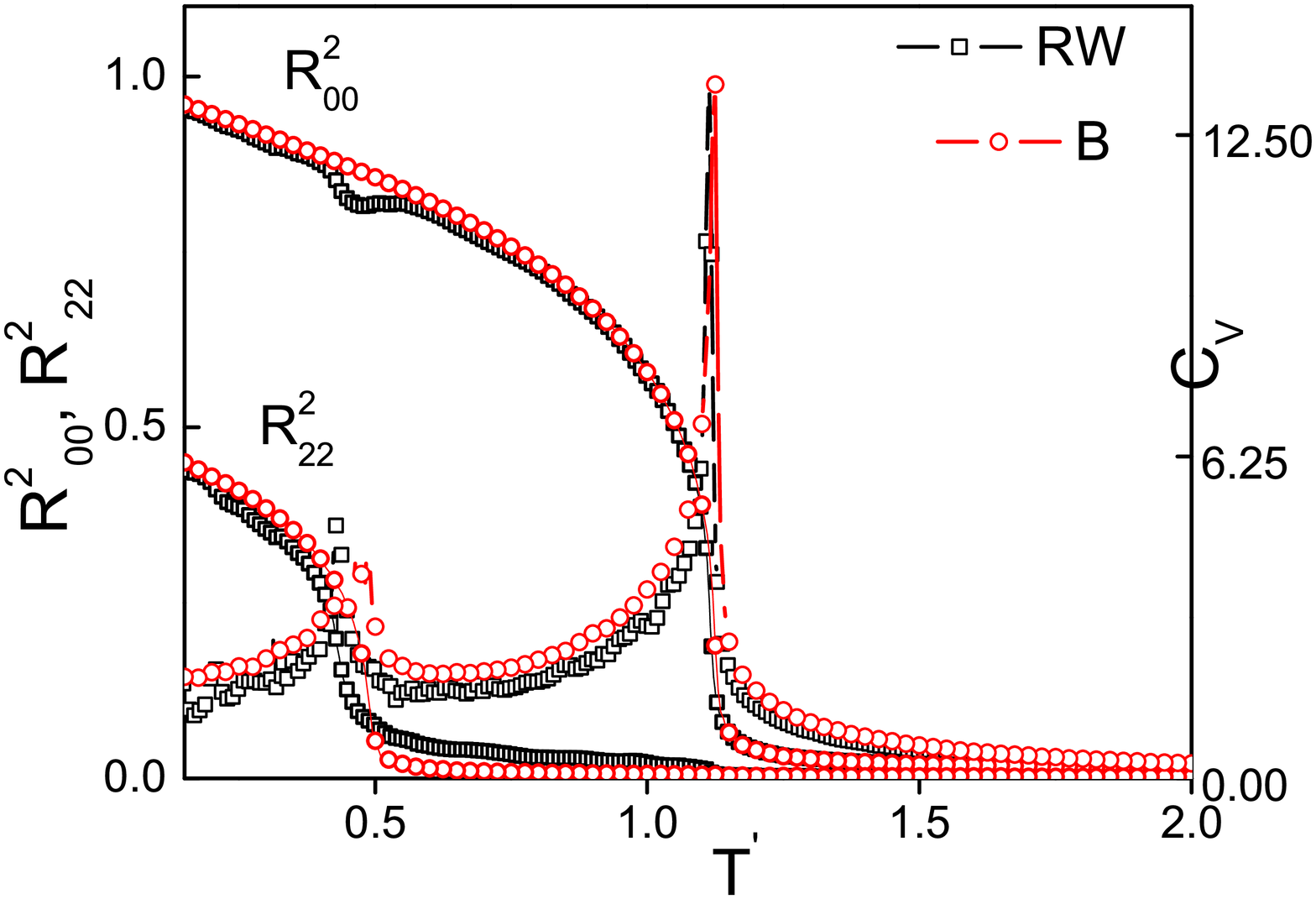}
\label{fig:6b}}
\subfigure[]{\includegraphics[width=0.35\textwidth]{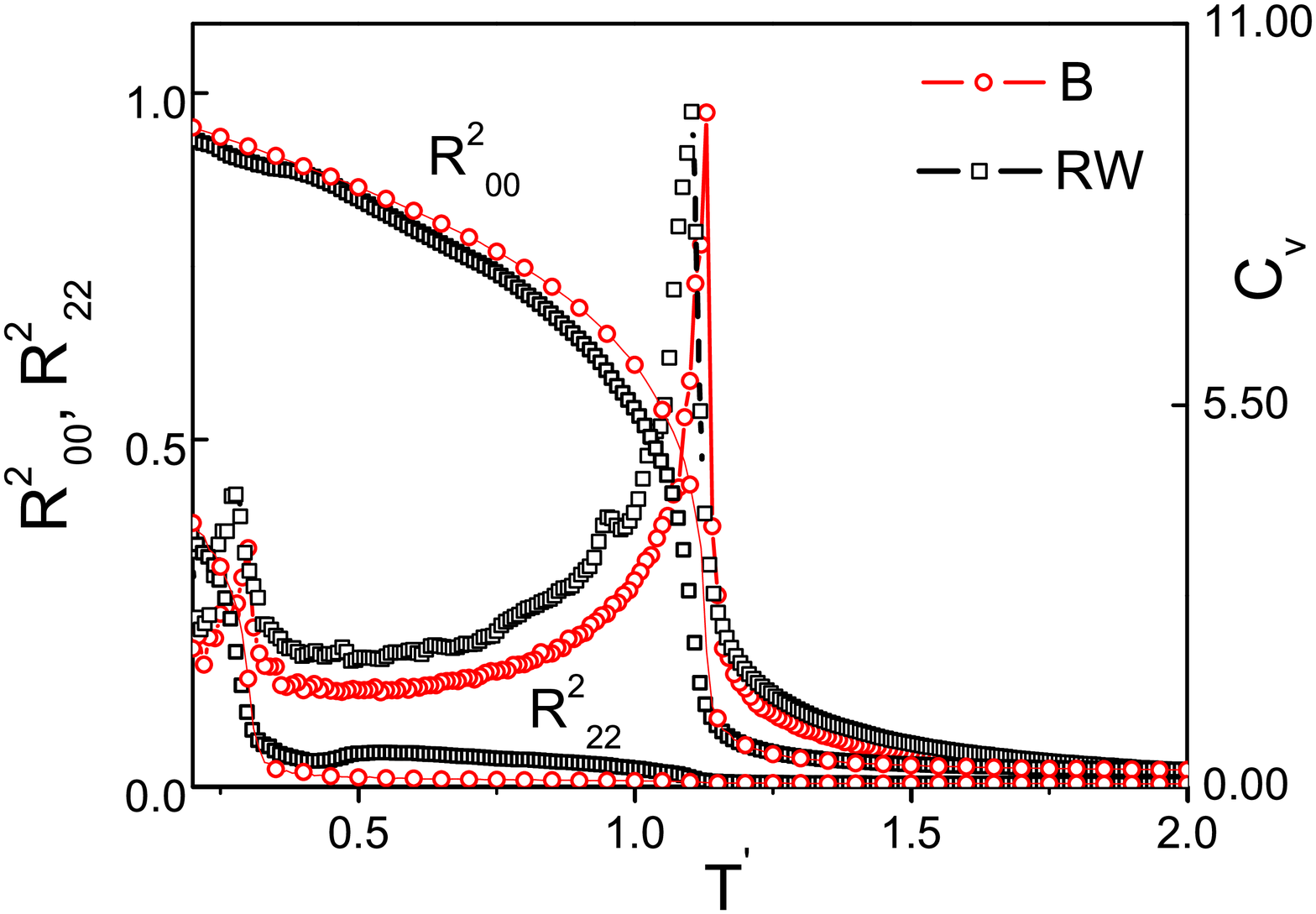}
\label{fig:6c}}
\subfigure[]{\includegraphics[width=0.35\textwidth]{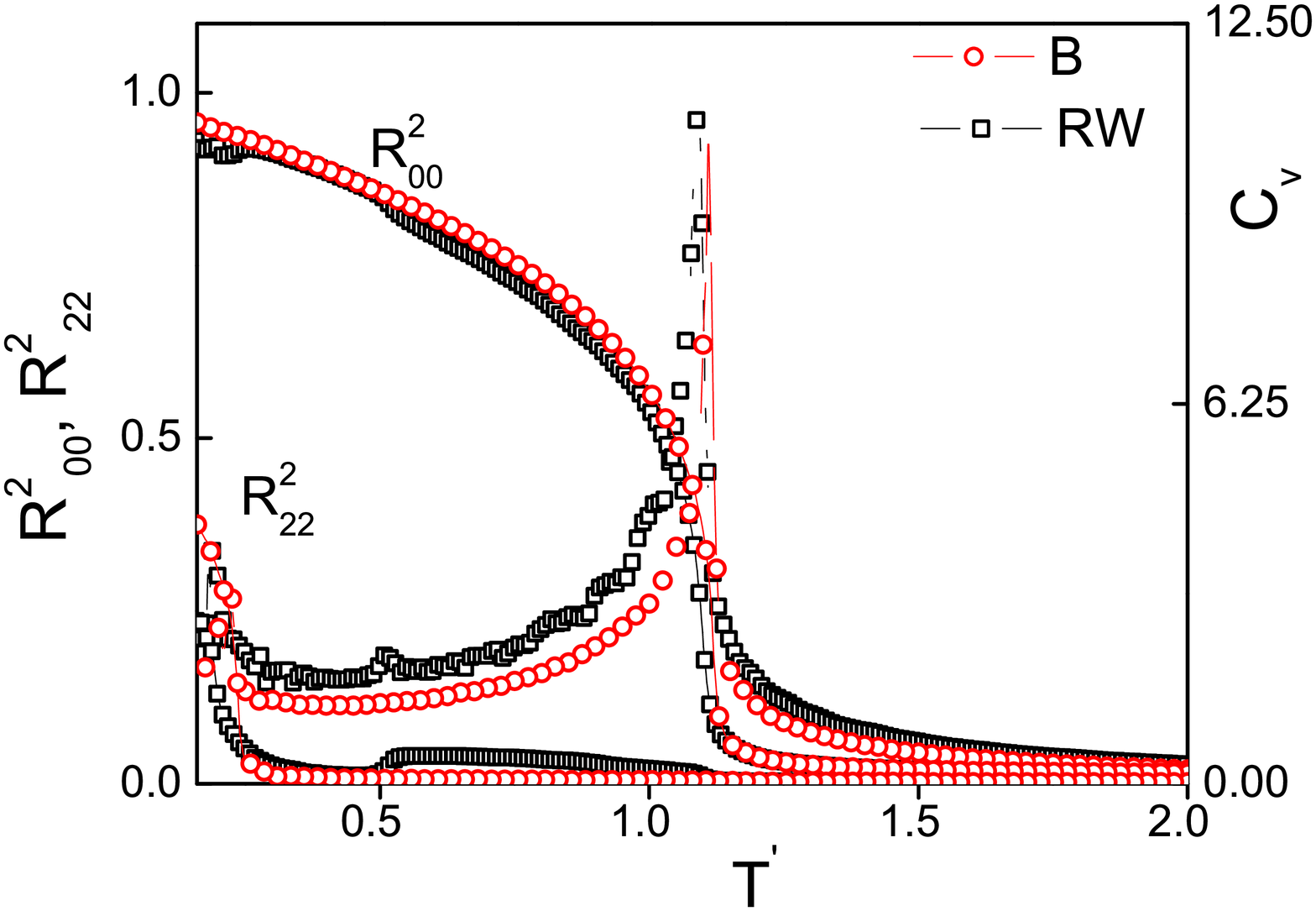}
\label{fig:6d}}
}
\caption{ (color online) Comparison of results obtained from RW ensembles
( hollow black squares) and B ensembles (hollow red circles). Temperature 
variation of the specific heat(continuous lines) and the order parameter 
profiles are shown for four values of $\lambda^{'}$ along the path $IW$:  
(a)0.566 (b) 0.656 (c) 0.692 (d) 0.709. It is seen that thermal averages of 
 $R^{2}_{22}$ from RW-ensembles  differ from the B-ensembles in the 
 intermediate $N_{U}$ phase for values of $\lambda^{'} > 0.566$. }
 \label{fig:6}
\end{figure}
We computed the equilibrium averages of the observables using B-ensembles
obtained from MC sampling at randomly chosen points on the trajectory IW.
A comparative study of the WL and MC simulation results at four such
representative  points (at $\lambda^{'}$ = 0.566, 0.656, 0.692 and 0.709) 
are shown in  Figs.~\ref{fig:6a} - \ref{fig:6d}.

 It is observed that qualitative agreement exists between the averages 
computed from RW- and B- ensembles upto (and including) $\lambda^{'}$ 
= 0.566. Thereafter the results vary in the behaviour of $R^{2}_{22}$ in the 
uniaxial phase. While the B-sampling results point to a pure uniaxial 
phase (i.e $R^{22}_{22} \sim 0 $ within the error bars) for  all 
values of $\lambda^{'}$ along the path IW, the RW-sampling results show 
an unmistakably non-zero and constant value of $R^{2}_{22}$ ($\sim$ 0.03) 
in the uniaxial phase for values of $ 0.566 < \lambda^{'} \leqslant 0.709$. 

\begin{figure}[t]
\centering
\includegraphics[width=0.8\textwidth]{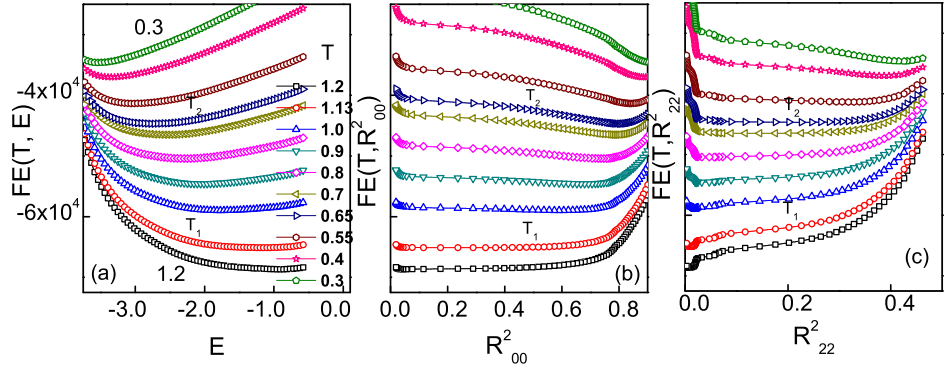}
\caption{(color online) Free energy shown as a function of 
(a) Energy per particle  E (b) $R^{2}_{00}$  and (c) $ R^{2}_{22}$, 
on cooling from the isotropic phase to the biaxial phase at 
$\lambda^{'}$ = 0.610.}
\label{fig:7}
\end{figure}
We show the representative free energy plotted as a function of 
energy (per particle) \textit{E} and the order parameters  $R^{2}_{00}$ and 
$R^{2}_{22}$ at $\lambda^{'}$ = 0.610 (point $B^{'}$) in the 
attractive region, in Fig.~\ref{fig:7}, at different 
temperatures bracketing the two transition points $T_{1}$ and $T_{2}$. We 
observe that the free energy minima with respect to energy shift towards 
lower values of energy, while
 they  shift towards higher values of order parameters progressively, 
as the system is cooled. We also note in Fig.~\ref{fig:7}(c), that the free
energy profile confines the value of $R^{2}_{22}$ to $\sim$ 0.03 
in the intermediate temperature region, before allowing  its access
to higher order values at the onset of $N_{B}$ phase at $T_{2}$. Further, 
the free energy profile thus confirms that the intermediate phase has to 
sport in principle a biaxial symmetry, though with a marginal value.

\begin{figure}
\centering
\includegraphics[width=0.4\textwidth]{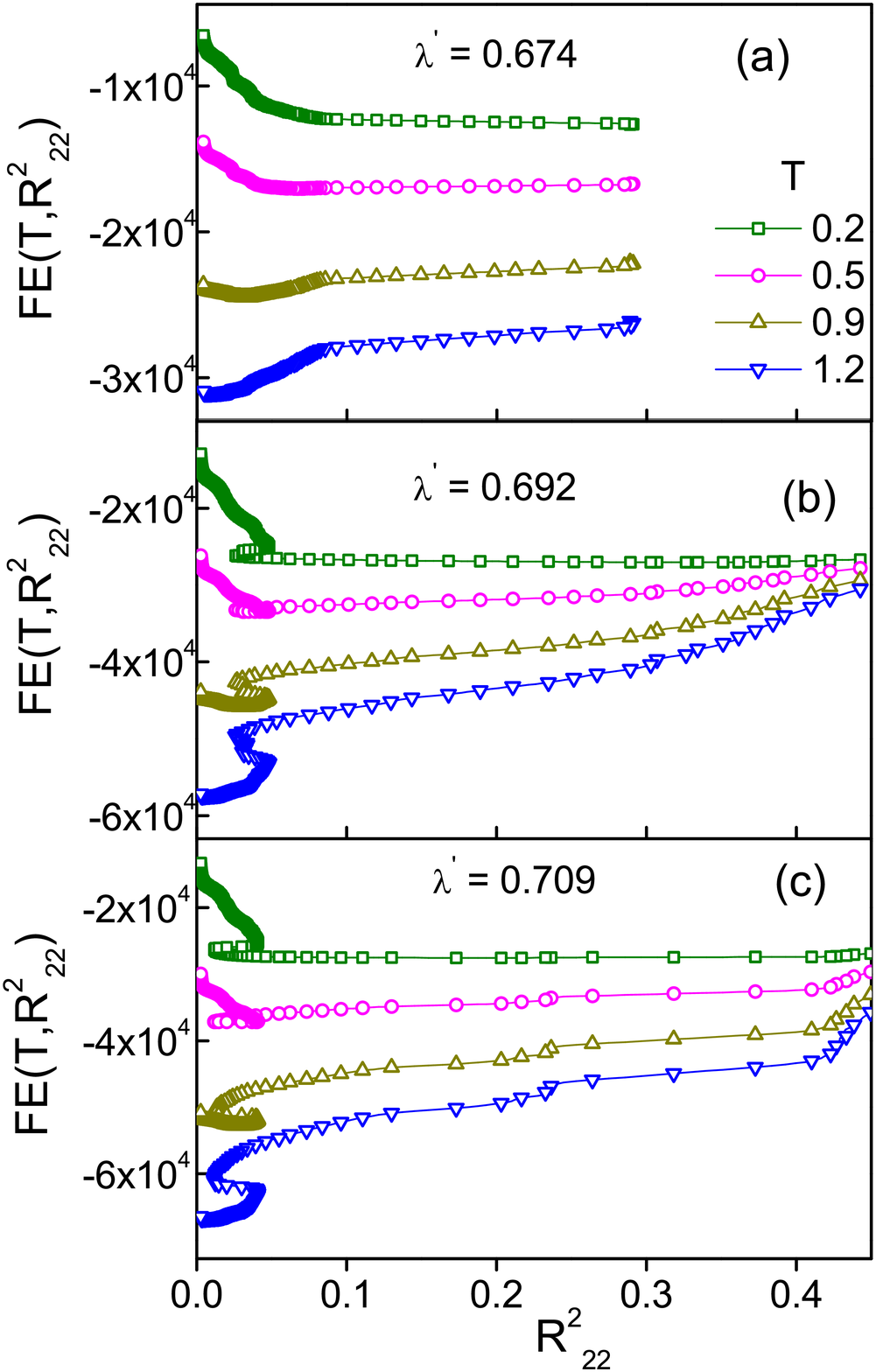}
\caption{(color online) Representative free energy plotted as a function of 
 $R^{2}_{22}$ for different values of $\lambda^{'}$ :  
 (a) 0.674 (b) 0.692  and (c) 0.709 (L=20 for (a) and (b) and L=15 
 for (c).}
\label{fig:8}
\end{figure}

However as we traverse from this fully attractive region of the
  Hamiltonian  towards the dispersion parabola bordering the repulsive
 region, the free energy profiles with respect the biaxial order 
 display curious deviations, and these persist
 on entering into the partly repulsive region of the Hamiltonian as well. 
 Fig.~\ref{fig:8} compares the temperature dependence of free energy 
 profiles plotted as a function of $R^{2}_{22}$,  at different points
 $B^{'}$, C, and D in the triangle (Fig.~\ref{fig:1}). Temperatures are 
 chosen to represent the profiles in different LC phases.
We note from Fig.~\ref{fig:8}(a) that the free energy curves at
 $\lambda^{'}$ = 0.674 (point $B^{'}$) show shallow minima  at finite 
values of $R^{2}_{22}$ (ranging from 0.03 in the $N_{U}$ phase to 
0.24 at the onset of the $N_{B}$ phase). In contrast free energy curves 
at $\lambda^{'}$ = 0.692 (Fig.~\ref{fig:8}(b) at point C located on the 
parabola) and 0.709 (Fig.~\ref{fig:8}(c) at point D, in the partly repulsive region) 
show a rather unusual behaviour in $N_{U}$ phase, however consistent with the 
temperature dependence of the observed average values of $R^{2}_{22}$, i.e 
initial increase to a higher value at the onset of $N_{U}$ phase and
subsequent dip on cooling just before the onset of $N_{B}$ phase.

 Noting the established accord between the temperature variation of
 average values of order parameters and the corresponding free energy
 profiles in this parameter region, and also keeping in view the observation
 that the free energy on the other hand shows a smooth variation with 
 the energy of the system, the obvious pointer is to suggest subtle 
 changes in the relative contributions of the different orders to the 
 entropy of the system. It seems to show rather pointedly that in the 
 neighbourhood of the parabolic boundary, the increased 
 contribution  of the cross-coupling term ($\gamma$), at
 the expense of the biaxial-biaxial coupling ($\lambda$) attempting to 
 promote macroscopic molecular biaxial order, do not leave the 
 intermediate unixial phase in its pristine form (compared to, say, 
 the nematic phase in LL-model or even in the biaxial system on the 
 $\lambda$-axis, for example). It may be noted that the presence of 
 such inhomogeneous  structures and their contribution to the macroscopic 
 averages of order have been investigated and the presence of 'clusters' 
 was alluded to, in the biaxial cluster model of nematics by Vanakaras
 \cite{Vanakaras08,Vanakaras09},  (which was proposed to explain the recent experimental 
 observation  of phase baixiality in bent-core nematics). Indeed this 
 specific uniaxial phase  seems to host local inhomogeneties catering 
 to increased $\gamma$ contribution and thus shows a non-zero macroscopic 
 $R^{2}_{22}$ initially originating from such clusters.  The subsequent decrease
 of biaxial order on further cooling in the uniaxial phase appears  
 to be an indication of  the increasing role of primary order parameter 
  $R^{2}_{00}$ in effectively contributing to the free energy minimization
  in the process  making the system perhaps a more homogeneous uniaxial medium. 
 
  \begin{figure}
\centering
\includegraphics[width=0.6\textwidth]{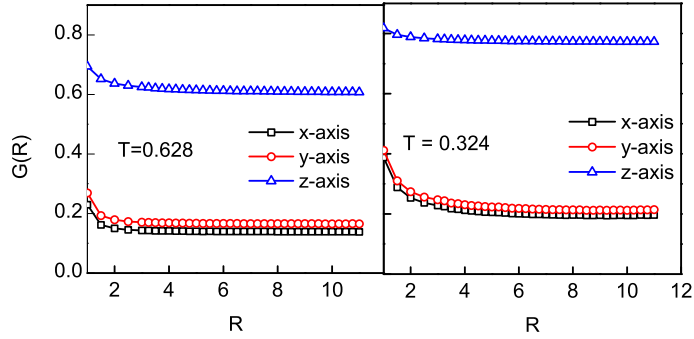}
\caption{(color online) Correlation function G(r) of x,y and z molecular 
axes plotted as a function of the distance r at the point C 
($\lambda^{'}$ = 0.692 ) at two temperatures in the $N_{U}$ phase. } 
\label{fig:9}
\end{figure} 
      Taking advantage of the non-monotonic variation of $R^{2}_{22}$
 within the $N_{U}$ phase at the point  C (on the parabola), we chose two
 temperature points (0.628 and 0.324) at which the average value of 
 $R^{2}_{22}$ are the same (Fig.~\ref{fig:2c}). We collected microstates 
 within a narrow range centered at the corresponding average energy
 values per site (-2.331 $\pm$ 0.001, -2.784 $\pm$0.001 respectively), constituting 
 effectively microcanonical ensembles located at the most probable 
 energy values at the respective temperatures. We computed the 
 orientational correlations of different molecular axes with distance
 (in lattice units), to obtain their spatial correlation functions
  at the two temperatures. These variations are shown 
 in Fig.~\ref{fig:9}. Obviously $R^{2}_{00}$ has increased significantly
 over this temperature range and is reflected in the long-range correlation 
 values of the z-axes. The minor axes (x and y) however have qualitatively
 different decays, flattening to two different plateau values, even though 
 the corresponding  macroscopic averages of $R^{2}_{22}$ are chosen equal.
  This clearly brings out the subtle differences in the microscopic 
  organization in the two  biaxial phases at the two temperatures: the 
  low temperature phase hosts a higher long-range $R^{2}_{00}$ order 
  as expected, but interestingly also a relatively  higher long-range 
  $R^{2}_{22}$  order. It may also be seen from the initial decay 
  profiles of the minor axes at the low temperature (Fig.~\ref{fig:9}),
  that this  hosts biaxial clusters which are correlated
  over larger range than their counterparts at the high temperature point. 
  The low temperature phase seems to correspond to an emerging
  homogeneous biaxial phase, homogeneity being perhaps imposed through 
  free energy considerations, by the inherent degree of the dominant
   uniaxial order $R^{2}_{00}$. The fact that these two temperatures 
 had the same macroscopic $R^{2}_{22}$ order, despite having 
 qualitatively differing correlation profiles, also confirm the presence 
 of contributions to $R^{2}_{22}$  possibly arising from geometrical 
 averages over inhomogeneous regions, at high temperature.
  
\subsection{ Study along the parabolic path OT }
  The parabolic path within the triangle extends from the origin \textit{O}
($\gamma$ = 0, $\lambda$ = 0, corresponding to the LL model) to the Landau 
point \textit{T} ($\gamma = 1/3$, $\lambda = 1/9$)
and the interaction parameters are related within the dispersion approximation 
as  $\lambda = \gamma^{2}$. We carried out entropic
sampling based MC study at 13 closely spaced points on the parabola 
(excluding the origin) and  the observed that phase sequence remained the same
$I - N_{U} - N_{B}$ at all points except at \textit{T}. The Landau point
 was found to be qualitatively different, hosting two distinct biaxial 
phases instead, as reported in a recent entropic sampling based MC study
\cite{Kamala14,Kamala15}. It may be noted that this finding however differs
 from the mean field prediction \cite{Freiser, Alben,Bisi06}
as well as  MC reults from Boltzmann sampling \cite{Biscarini95}.
The latter studies point to a single low temperature $N_{B}$ phase after a 
direct transition from the isotropic phase.
\begin{figure}
\centering
\includegraphics[width=0.5\textwidth]{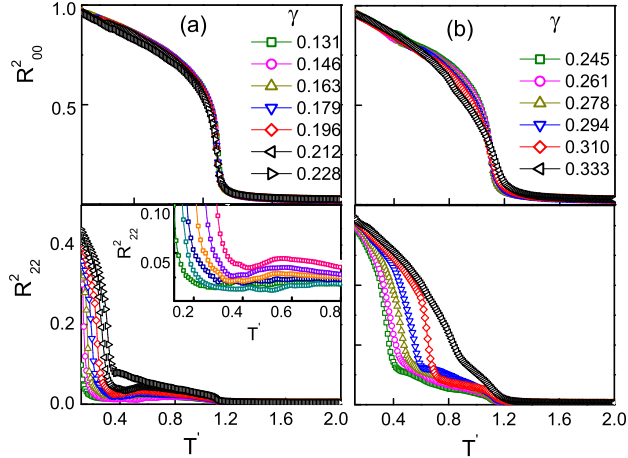}
\caption{(color online) Uniaxial order ($R^{2}_{00}$) and biaxial order
($R^{2}_{22}$) plotted as a function of reduced temperature for values 
of $\gamma$ ranging from (a) 0.131 - 0.228 (b) 0.245 - 0.333. Inset in (a)
shows a magnified version of the $R^{2}_{22}$ vs $T$ plot where the 
decrease of  $R^{2}_{22}$ at lower temperatures is seen. } 
\label{fig:10}
\end{figure}    
 We present the order parameter profiles at various points on the 
 parabola starting from $\gamma$ = 0.131 to $\gamma = 1/3$
 in Fig.~\ref{fig:10} (at L = 20). It is observed from Fig.~\ref{fig:10}(a) that
 the biaxial order parameter shows an initial small increase at the 
 onset of the $I - N_{U}$ transition, followed by a decrease
 in its value in the deeper uniaxial nematic  phase. This anomalous 
 behaviour is more pronounced for values of $\gamma$
 ranging from 0.163 - 0.212. The $R^{2}_{22}$ temperature profiles  
 for $\gamma$ values in the range 0.245 - 0.333 on the other hand, 
  increase continuously in the $N_{U}$ phase (Fig.~\ref{fig:10}(b))
 exhibiting a monotonic behaviour. It is observed from both the graphs that
the temperature range of the uniaxial nematic phase 
decreases and biaxial phase appears at progressively higher
temperatures, as the $\gamma$ value increases along the parabola.
A curious observation from this study is that the intermediate $N_{U}$ 
phase is not strictly uniaxial with $R^{2}_{22}$ = 0 (as expected from
the earlier studies), but hosts a small degree of biaxial order in the 
intermediate temperature range. This feature becomes prominent as $\gamma$
value increases beyond $\sim$ 0.2, indicating the increasingly
competing role of the cross-coupling interaction on this very special boundary.
\begin{figure}
\centering
\includegraphics[width=0.5\textwidth]{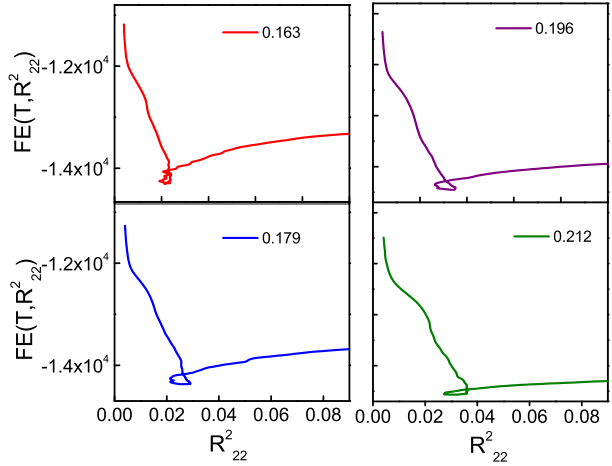}
\caption{(color online) Free energy plotted as a function the biaxial 
order parameter $R^{2}_{22}$ for four values of $\gamma$ in the 
neighbourhood of point C (including C, $\gamma$ = 0.212)} 
\label{fig:11}
\end{figure} 
         The free energy profiles (plotted against $R^{2}_{22}$) shown
 in Fig.~\ref{fig:11} at temperature T = 0.5 for values of $\gamma$
 between 0.163  and 0.212 on the parabola, show the  presence of loop 
 like structures, similar to the earlier observations at point C
 (Fig.~\ref{fig:8}(b)), and consistent with the temperature variation of
 average $R^{2}_{22}$ values. 
 \begin{figure}
\centering 
\includegraphics[width=0.4\textwidth]{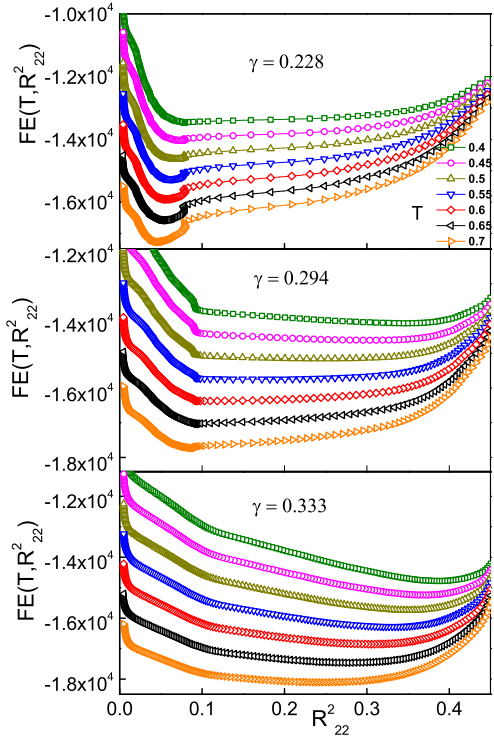}
\caption{(color online) Free energy shown as a function of 
 $ R^{2}_{22}$, on cooling in the uniaxial nematic phase for 
 values of $\gamma$ = (a) 0.228 (b) 0.294 (c) 0.333 }
\label{fig:12}
\end{figure}    
  For higher values of $\gamma$ (0.228 - 0.333,
  Fig.~\ref{fig:12}) however, these show variations
 on cooling, which are in accord with the behaviour of $R^{2}_{22}$ 
 in this region of the parabola.
 
      Thus it emerges, that the intermediate $N_{U}$ phase hosts distinct 
molecular organizations as the medium is transformed in terms of the 
symmetry of its  molecular interactions from LL - model to the Landau point
along the parabola. Discernible degree of biaxial order and its curious 
temperature variations along the path starting from the origin (LL model)
hint at the possibility that the parabola is in fact a very special trajectory 
having differing types of $N_{U}$ phases as the Landau point is reached. 
The parabola at once serves both as an interesting 
boundary between distinct natures of the Hamiltonian, as well as transforms
the system interaction symmetry while simultaneously promoting the influence
of the cross-coupling terms, as one moves over from the LL limit.
Obviously non-monotonic temperature dependence of $R^{2}_{22}$ is associated
with complex free energy terrain exhibited by the system in the 
$R^{2}_{00}$ - $R^{2}_{22}$ space, originating from increasing degree of 
cross-coupling term. Viewed from this perspective, the present 
data provide an insight into the role of $\gamma$ and $\lambda$ 
as their relative importance changes on this trajectory.
 
\section{Conclusions }
We report the results of detailed MC simulations (based on Wang - Landau
technique) along two trajectories inside the triangle $\Delta$.
In the first case, along the line IW, we find that our results 
are in accord with MF predictions in terms of the phase sequences
expected. We observe however that as we progressively move towards the
base point W, in the process changing the relative importance of 
$\gamma$ and $\lambda$ terms in Eqn.~\ref{eqn:w1}, the uniaxial
phase develops a marginal degree of biaxial order $R^{2}_{22}$ which
is sustained through the uniaxial range. This is similar to our earlier
observation on the diagonal IV as one progressively traverses towards V
\cite{Kamala14,Kamala15}. The onset of a biaxial phase with significant
order at the low temperature transition is preceded by a dip in 
$R^{2}_{22}$ from its small value ($\sim$ 0.03).

     The trajectory IW encompasses two distinct regions from the point of 
view of nature of the Hamiltonian. Upto the point C where IW intersects
the dispersion parabola, \textit{H} is fully attractive. The segment 
CW corresponds to a partly repulsive region, making the stability of 
the biaxial phase untenable asymptotically as the point W is reached. 
We make use of the DoS estimates in our simulation to plot the free energy
profiles as function of order parameters ($R^{2}_{00}$ , $R^{2}_{22}$)
as well as energy. The observed interesting temperature variation of 
$R^{2}_{22}$  within the 'uniaxial' phase is consistent  with the
requirements of the free energy profiles at different $\lambda^{'}$
values along the trajectory.   

     The parabolic trajectory \textit{OCT}, very well studied earlier
 for its simplifying dispersion approximation, is revisited with the 
 present MC technique to examine if the intermediate uniaxial phase 
retains its pristine symmetry ($R^{2}_{22}$ = 0 in this phase)
throughout its path. The present data indicate that the intermediate
uniaxial phase exhibits a small degree of biaxial order as $\gamma$
increases, and as Landau point is reached it indeed seems to transform
into a biaxial phase in its own right \cite{Kamala14, Kamala15}.

   The appearance of a small degree of biaxial symmetry within the 
uniaxial phase, whenever $\gamma$ assumes a dominant role, has its 
origin in the presence of local biaxial inhomogeneities (referred
to as 'clusters' in \cite{Vanakaras08,Vanakaras09}). Their formation 
and sustenance is facilitated by the corresponding cross-coupling 
interaction which eventually interferes with the homogeneous onset of 
the two orders. This inference may well have implications in the 
observed difficulties in realizing readily a biaxial phase in the 
laboratory.  

\section{Acknowledgments}
We wish to thank Surajit Dhara, School of Physics, University of
 Hyderabad for useful discussions.  The simulations were carried out in 
 the Centre for  Modelling Simulation and Design (CMSD) at the 
 University of Hyderabad. BKL acknowledges financial support
 from Department of Science and Technology, Government of India  vide 
 grant ref No: SR/WOS-A/PM-2/2016 (WSS) to carry out this work.

\end{document}